\documentclass{article}
\usepackage{spconf,amsmath,amssymb,graphicx,mathtools, nccmath}
\renewcommand{\paragraph}[1]{\noindent\textbf{#1}\quad}
\usepackage{caption, subcaption}
\usepackage{multirow}
\usepackage{float}
\usepackage{lipsum}
\usepackage{hyperref}
\usepackage{booktabs}
\usepackage{mathtools}
\usepackage{etoolbox}
\usepackage{xcolor}
\usepackage{xpatch}
\makeatletter
\g@addto@macro\normalsize{%
	\setlength\abovedisplayskip{4pt}
	\setlength\belowdisplayskip{4pt}
	\setlength\abovedisplayshortskip{4pt}
	\setlength\belowdisplayshortskip{4pt}
}

\DeclareMathOperator*{\argmax}{arg\,max}

\usepackage[capitalize]{cleveref}
\Crefname{chapter}{Chap.}{Chaps.}
\Crefname{section}{Sec.}{Secs.}
\Crefname{figure}{Fig.}{Figs.}
\Crefname{table}{Table}{Tables.}

\usepackage{pifont}
\name{Author(s) Name(s)\thanks{Thanks to XYZ agency for funding.}}
\address{Author Affiliation(s)}

\title{Improving Factored Hybrid HMM Acoustic Modeling without State Tying}
\name{Tina Raissi$^1$, Eugen Beck$^{2}$, Ralf Schl\"uter$^{1,2}$, Hermann Ney$^{1,2}$ }

\address{\vspace{-0.1cm}
	$^1$Human Language Technology and Pattern Recognition Group, RWTH Aachen University, Germany \\ $^2$AppTek GmbH, Aachen, Germany \\\vspace{-0.1cm} \textit{\{raissi,schlueter,ney\}@cs.rwth-aachen.de, ebeck@apptek.com}}

\begin{document}
%
\maketitle
\begin{abstract}
In this work, we show that a factored hybrid hidden Markov model (FH-HMM) which is defined without any phonetic state-tying outperforms a state-of-the-art hybrid HMM.\ The factored hybrid HMM provides a link to transducer models in the way it models phonetic (label) context while preserving the strict separation of acoustic and language model of the hybrid HMM approach. Furthermore, we show that the factored hybrid model can be trained from scratch without using phonetic state-tying in any of the training steps.\ Our modeling approach enables triphone context while avoiding phonetic state-tying by a decomposition into locally normalized factored posteriors for monophones/HMM states in phoneme context. Experimental results are provided for Switchboard 300h and LibriSpeech.\ On the former task we also show that by avoiding the phonetic state-tying step, the factored hybrid can take better advantage of regularization techniques during training, compared to the standard hybrid HMM with phonetic state-tying based on classification and regression trees (CART).

\end{abstract}

\vspace{-0.1cm}
\section{Introduction}
\label{sec:intro}
\vspace{-0.3cm}
In automatic speech recognition~(ASR) with the hybrid neural network/hidden Markov model~(NN/HMM) approach~\cite{bourlard1994hybrid}, the co-articulation effect is addressed by considering tied allophone states in right and left context, known also as senones.\ The parameter tying relies generally on the state clustering done by classification and regression trees~(CART)~\cite{odell1994tree}, and requires a frame-wise alignment of the input speech signal to a hidden HMM state sequence.\ This alignment, can be obtained from a Gaussian Mixture Model~(GMM) system, and is used for both estimation of CART and frame-wise cross-entropy~(CE) training of the neural network.\ The NN component is used for the substitution of the likelihood of a given feature vector by the normalized scaled posterior probability over the tied state label.\ The overall system has separate components and requires different sources of knowledge.\ Moreover, having tied state labels as the NN target can lead to overfitting problems due to the arbitrary choices done for the estimation of the clustering~\cite{bell2016multitask}.\

In order to simplify the described pipeline, the recent research focus has shifted to the joint optimization of the different components in a sequence to sequence~(seq2seq) framework.\ In this approach, the speech sequence is directly mapped to the final word sequence, with an optional use of an external LM.\ In connectionist temporal classification~(CTC)~\cite{ctc}, the necessity of having an initial frame-wise alignment is relaxed by introducing a blank label and using a sequence-level CE which sums over all possible alignments.\ The CTC criterion contains an independence assumption at each time frame.\ The inclusion of the dependency to the full or truncated context is introduced with Recurrent Neural Network Transducers~(RNN-T)~\cite{rnnt} and Recurrent Neural Aligner~\cite{rnaSak} with different constraints in how to traverse the alignment lattice.\ Other prior work focuses on the elimination of state-tying for both hybrid HMM and CTC frameworks using full set of diphone states~\cite{povey} or a triphone representation via an embedding network~\cite{chorowski2019towards}.\ The aforementioned seq2seq models, are similar to the hybrid approach in providing a statistical model of the posterior distribution over different label topologies and units.\ This posterior is later used in a (weighted) finite state transducer like structure for decoding.\ However, they do not use a local normalization and the correct way to combine an external LM is still an open question~\cite{ilm1,variani2020hybrid,zeineldeen2021investigating}.\ The modular formulation in the hybrid approach results in an intrinsic robustness for tasks such as domain adaptation and for low resource ASR tasks~\cite{pawel}.\ Furthermore, seq2seq models require large amounts of data and many additional regularization techniques in order to reach a reasonable performance.\ Nevertheless, hybrid models with alternative label units and encoder architectures~\cite{facebook,transformer} can still have competitive performance compared to highly optimized seq2seq pipelines~\cite{tuskernnt,zoltan}.\ We believe that the first step for a possible simplification of the standard hybrid pipeline is the elimination of the state-tying algorithm.\ Recently, we presented a factored hybrid HMM model~\cite{raissi}.\ Given the complete set of triphone states, the factored model learns to predict context-dependent posteriors over phonemes, in a multi-task manner.\ This is achieved by a decomposition of the joint label identity of the center phoneme state and its left-right phonemes.\ With this previous work, and its extension~\cite{raissi2021towards}, we have shown that it is possible to eliminate the state-tying for the determination of the state inventory, and obtain similar performance to a hybrid CART.\

Our main contribution with this work is three-fold: (1) We eliminate the dependency to the clustering algorithm entirely by proposing a from scratch pipeline.\ With that we obtain slightly better performance compared to the standard hybrid using state-tying, (2) We show that our factored hybrid outperforms the standard approach when using the same initial alignment. And, (3) we show that our modeling approach can obtain larger improvement from regularization techniques, also on a small task such as 300 hours Switchboard~(SWB), compared to the CART based hybrid.\ We also present a set of baseline results using LibriSpeech~(LBS), without the main regularization techniques, confirming the scalability of our approach to a larger task.\
 
\section{Modeling Approach}  
\label{sec:model}

We rely on our previously presented approach~\cite{raissi}, with the set $\left\lbrace \phi_{\ell},\sigma_c,\phi_{r}\right\rbrace_t$ of left phoneme, center phoneme state and right phoneme of the aligned triphone at time frame $t$.\ We substitute the emission probability of the feature vector $x$, resulting from the decomposition of the acoustic component as follows: 
\begin{equation}\footnotesize 
p_t(x|\phi_{\ell},\sigma_{c},\phi_{r})=\frac{p_t(\phi_{\ell},\sigma_{c},\phi_{r} | x)\cdot p_t(x)}{p(\phi_{\ell},\sigma_{c},\phi_{r})} \nonumber
\end{equation}
The idea is to carry out the factorization into the left-to-right trigram of \cref{eq:fwd}, with additional diphone and monophone formulations, as shown in \cref{eq:di} and \cref{eq:mono}, respectively.  
\begin{subequations}
\begin{align}\footnotesize
p_t(x | \phi_{\ell},\sigma_{c},\phi_{r})&\sim \frac{p_t(\phi_{r}|\phi_{\ell},\sigma_c, x)\cdot p_t(\sigma_c|\phi_{\ell}, x) \cdot p_t(\phi_{\ell}|x)}{p(\phi_{r} | \phi_{\ell},\sigma_c)\cdot p(\sigma_c|\phi_{\ell}) \cdot p(\phi_{\ell})}\label{eq:fwd} \\
p_t(x | \sigma_c,\phi_{\ell}) &\sim \frac{p_t(\sigma_c|\phi_{\ell}, x)p_t(\phi_{\ell}|x)}{p(\sigma_c|\phi_{\ell}) p(\phi_{\ell})}\label{eq:di} \\
p_t(x | \sigma_c) &\sim \frac{p_t(\sigma_c|x)}{p(\sigma_c)} \label{eq:mono}
\end{align}
\end{subequations}
In practice, the decision rule in the maximum a-posteriori framework, with input feature $x_1^T$, and word sequence $w_1^N$, is done via log-linear combination of different factors.\
\begin{align} \label{eq:bayes}\footnotesize
\underset{w_1^N}{\argmax} & \left\lbrace  \max_{\{\phi_{\ell},\sigma_c,\phi_{r}\}_1^T:w_1^N}  \left. \Bigg[ \sum_{t=1}^{T} \log p_t(\phi_{r}|\sigma_c, \phi_{\ell}, x)  \right.\right. \\ \nonumber
&- \left. \left. \gamma_r \cdot \log p(\phi_{r}|\sigma_c, \phi_{\ell}) +  \log p_t(\sigma_c|\phi_{\ell}, x) \right. \right.\\ \nonumber
&-\left. \left. \gamma_c \cdot \log p(\sigma_c|\phi_{\ell})+ \log p_t(\phi_{\ell} | x)\right.\right.\\\nonumber
&-\left. \left.  \gamma_{\ell} \cdot \log p(\phi_{\ell}) +  \beta \cdot \mathcal{T}(t,t-1)  \Bigg] \right. \right.\\\nonumber
&+\left.  \alpha \cdot \sum_{n=1}^{N}\log p(w_n|w_{n-3}^{n-1})  \right\rbrace\nonumber
\end{align}

\noindent Our general formulation is shown in \cref{eq:bayes} using prior, transition penalty, and language model~(LM) scales $\gamma$, $\beta$, and $\alpha$, respectively.\ Here, we consider a 4-gram LM.\ The term $\mathcal{T}$ is a constant representing the state transition probabilities, with distinct values in case of speech or silence.\  
\vspace{-0.2cm}
\section{Training and Decoding}
\label{sec:train}
\vspace{-0.2cm}
The common starting point for all experiments is the initialization of the parameters of a context-independent single Gaussian HMM based on a linear segmentation of the input signal.\ Using the resulting alignment, a GMM monophone system is bootstrapped.\ We denote the resulting alignment as \textbf{GMM-Mono}.\ For the CART system, the set of context-dependent tied states are first clustered in order to build a GMM triphone system.\ After the inclusion of the speaker adaptation, another CART for clustering the final HMM state inventory is estimated.\ Each described step consists of different iterations of parameter re-estimation and subsequent re-alignments.\ We denote the intermediate alignment obtained from the speaker-adapted triphone system as \textbf{GMM-Tri}.\ For the final best alignment, which we denote as \textbf{Tandem}, first a tandem model with speaker adaptation is trained, followed by miniumum phone error sequence discriminative training~\cite{tuske2015asru}.\ All the described steps above are necessary for the CART based system in order to reach a good performance.\ For each of the alignments described above, we compare Word Error Rates~(WERs) of our factored hybrid models.\ We use GMM-Mono and Tandem alignments for Switchboard 300h, and GMM-\{Mono,Triphone\} alignments for Librispeech 960h.\ 

The triphone factored hybrid, as described in \cref{eq:fwd} has three separate softmax layers: two of them have as target the left and right phonemes and the third is over the center phoneme states.\ The context-dependency for the right phoneme and  center state is fed via a soft embedding.\ The overall training procedure consists of multiple stages.\ In each stage, we increase the model complexity as follows.\ We start with a monophone model without feeding any context.\ In the following stages, we increase the context-dependency by feeding the context via phoneme and phoneme state embeddings~\cite{raissi}.\ The effect of the multi-stage training is reported in \cref{tab:pretrain}.\ 

In addition to the common regularization techniques described later in \ref{subsec:setting}, we focused on the effect of the following two techniques.\ \textbf{Label Smoothing~(LS)} is usually used in order to contrast the model's overconfidence about the aligned target~\cite{labelsmooth}.\ The minimization of the cross-entropy in LS is done with respect to a soft target.\ The target is obtained by interpolating a uniform distribution over other classes with the aligned target.\ Differently to the monophone targets, where the problem is well-defined, the application of LS is less straightforward when context-dependent targets are considered.\ Since the posterior over the untied triphone state corresponds to the combination of the factored posteriors, the idea is to keep a hard target for the center phoneme state and apply LS only on the context.\ Consequently, starting from the diphone stage we turn off LS on the center state output.\ Experimental results, as reported in \cref{tab:ls}, for a CART-free pipeline starting with the GMM-Mono alignment confirm the following: (1) The relative improvement obtained from LS in monophone model is larger than in di- and triphone models, (2) continuing with LS on all outputs leads to degradation for diphone model.\ For CART based models application of LS did not help.\ In this case, each label corresponds to a set of tied triphone states, and therefore the model learns to discriminate between state classes defined by the clustering and not by the triphone state identity itself.\ We also tried \textbf{Specaugment}, based on random on-the-fly time and feature masking used widely in seq2seq training~\cite{park2019specaugment}.\ All models are trained with the same masking parameters, following indications from~\cite{weimask}.\
\vspace{-0.2cm}
\subsection{Experimental Setting}
\label{subsec:setting}\vspace{-0.3cm} 
    \begin{table}[t]    	
    	\setlength{\tabcolsep}{0.8em}\renewcommand{\arraystretch}{1.1}  
    	\centering \footnotesize
    	\caption{Comparison of the performance of factored triphone models traind on SWB 300h initialized with random parameters and initialized with a diphone model, with no label smoothing, no data augmentation and no realignment.\ Evaluation is on Hub5'00 with 4-gram LM.\ }
    	\label{tab:pretrain} \vspace{-0.2cm}
    	\begin{tabular}{|c|c|c|c|} 
    		\hline     		
    		\textbf{Model} &\textbf{Alignment} &\textbf{Pretrained} &\textbf{WER[\%]} \\ \hline
    		CART     & \multirow{3}{*}{Tandem} & \multirow{2}{*}{no} & 13.4 \\\cline{1-1} \cline{4-4}
    		\multirow{3}{*}{FH} &     &                      & 13.7 \\ \cline{3-4}
    		& & yes & 13.5 \\ \cline{2-4}
    		\multirow{2}{*}{Triphone} & \multirow{2}{*}{GMM-Mono}     &no        & 15.2 \\ \cline{3-4}
    		& & yes & 14.5 \\ \hline   		
    	\end{tabular}
    	\vspace{-0.2cm}
    \end{table}   	
 \begin{table}[] 	
 	\setlength{\tabcolsep}{0.6em}\renewcommand{\arraystretch}{1.2}  
 	\centering \footnotesize
 	\caption{Label Smoothing in SWB 300h for targets in the output layers of CART based and different factored hybrid~(FH) models, evaluated on Hub5'00 with 4-gram LM.\ Rows with the same color correspond to one multi-stage pipeline, meaning that n-phone model is initialized with (n-1)-phone model.\ }\vspace{-0.2cm}    	
 	\label{tab:ls}
 	\begin{tabular}{|c|c|c|c|c|c|} 
 		\hline
 		\multirow{2}{*}{\textbf{Model}} & \textbf{N-} &  \multicolumn{3}{c|}{\textbf{Label Smoothing on Target}}& \multirow{2}{*}{\textbf{WER[\%]}} \\ \cline{3-5}
 		& \textbf{phone}&\multicolumn{3}{c|}{\textbf{Tied Triphone State}}& \\ \hline \hline
 		
 		\multirow{2}{*}{\textbf{CART}}& \multirow{2}{*}{Tri}& \multicolumn{3}{c|}{no} & \textbf{13.4}\\ \cline{3-6}
 		&                     & \multicolumn{3}{c|}{yes}& 13.7 \\ \hline 
 		\multicolumn{2}{|c|}{}   &\textbf{Left}&\textbf{Center}&\textbf{Right}& \multicolumn{1}{c|}{}\\ \hline 
 		\multirow{7}{*}{\textbf{FH}}&	\multirow{2}{*}{Mono}& \multicolumn{3}{c|}{\color{blue}{no}}& 17.1\\ \cline{3-6} 
 		&&\multicolumn{3}{c|}{\color{purple}{yes}}&\textbf{16.3} \\ \cline{2-6} 
 		\multirow{4}{*}{\textbf{}}&	\multirow{3}{*}{Di}&\multicolumn{3}{c|}{\color{blue}{no}}& 15.2\\  \cline{3-6} 
 		&&\multicolumn{3}{c|}{yes$^\star$}& 15.4\\ \cline{3-6} 
 		&&\color{purple}{yes}&\color{purple}{no}&\color{purple}{yes}& \textbf{14.9}\\ \cline{2-2} \cline{3-6} 
 		&\multirow{2}{*}{Tri}&\multicolumn{3}{c|}{\color{blue}{no}}& 14.5\\ \cline{3-6} 
 		&&\color{purple}{yes}&\color{purple}{no}&\color{purple}{yes}&\textbf{14.4} \\ \hline
 	\end{tabular}
 	\\$\star$ Also initialized with monophone using label smoothing on all targets. \vspace{-0.5cm}
 \end{table}

The experiments are conducted on 300h Switchboard-1~(SWB) Release 2~(LDC97S62)~\cite{godfrey1992switchboard} and 960 hours Librispeech~(LBS) \cite{panayotov2015librispeech} tasks, by using RETURNN and RASR toolkits~\cite{zeyer2018returnn,wiesler2014rasr}\footnote{\scriptsize {See code: https://github.com/rwth-i6/returnn-experiments/tree/master/2022-swb-factored-hybrid.}}.\ The evaluations for the SWB task are done on SWB and CallHome~(CH) subsets of Hub5'00 (LDC2002S09) and three SWB subsets of Hub5'01 (LDC2002S13).\ For LBS we report word error rates on both dev and test sets.\ The baseline CART system for Librispeech follows the system described in~\cite{luscher2019rwth}.\ The SWB systems for both CART and factored hybrid share the same setting and similar model sizes.\ We kept the same number of parameters for the BLSTM encoder for all experiments, comprising 6 forward and backward layers of size 512 with $10\%$ dropout probability~\cite{srivastava2014dropout}.\ The input speech signal to the encoder is represented by (SWB: 40 and LSB: 50) dimensional Gammatone Filterbank features~\cite{schluter2007gammatone}, extracted from 25 milliseconds~(ms) analysis frames with 10ms shift.\ Concerning the factored architectures, the one-hot encoding of the left and right phonemes, and the center phoneme states are projected by using linear layers of dimension 64 and 256, respectively.\ A two-layer perceptron in all factored models are used to combine the embedding and encoder outputs.\ The state inventory of both corpora consists of the complete set of triphone states, corresponding to a tripartite 0-1-2 HMM label topology for each phoneme in context, and a special context-independent silence state.\ Differently to the standard LSB phoneme inventory, we removed stress markers from all phonemes.\ For the standard hybrid, a set of (SWB:9001, LSB:12001) CART labels are considered.\ All models are trained with frame-wise CE using external alignments presented earlier in \cref{sec:train}.\ In order to speed-up the training, the sequences are divided into chunks of length $128$ with $50\%$ overlap, and fed to the network in batches of 10k.\ Our models on all tasks are trained using an Adam optimizer with Nesterov momentum~\cite{dozat2016incorporating}.\ We also use Newbob learning rate~(LR) scheduling with an initial LR of $1e^{-3}$, and a decay factor of $0.9$ based on the frame error rates.\ The minimum LR  value is set to (SWB:$2e^{-5}$, LSB:$1e^{-5}$).\ For the factored hybrid the reference error value results from the summation of the frame errors from the three outputs.\ In addition to label smoothing with scale factor $0.2$, we use also other regularization techniques: $L_2$ weight decay with a scale of $0.01$, gradient noise~\cite{neelakantan2015adding} with a variance of $0.1$, and focal loss factor of $2.0$~\cite{lin2017focal}.\ CART model converges after (SWB: 24, LSB: 6.25) epochs.\ Factored models are generally trained up to maximum 30 epochs for each stage.\ The optimal tradeoff between number of epochs in combination with multi-stage phonetic training has still not been explored and is left for future work.\ For recognition we use 4-gram~\cite{4gram} and LSTM language models~\cite{lstm}.\ We also include a second pass rescoring with a Transformer~(Trafo) LM for one of our experiments~\cite{irie2019training}.\ In addition to the LM scale, the separate transition penalty and prior scales of \cref{eq:bayes} are tuned.\ The explicit context modeling and the decomposition into context-dependent sub-architectures of the factored model require a careful implementation for the computation of the acoustic scores.\ The search space organization in case of both models follows the lexical prefix tree structure, where by means of dynamic programming, a time-synchronous beam decoding is carried out~\cite{ney1992improvements,ney1999dynamic}.\ At each time frame a set of path hypotheses are propagated and pruned to the most promising ones.\ Thus, not all the left-center context pairs are actually needed at each step.\ More specifically, we only need to forward in one batch the set of active left-center context pairs and cache the scores.\ By taking advantage of this aspect, we reduced the RTF from $12.6$ to $5.3$, i.\ e.\ by $59\%$ relative.\ For a similar search space size, the RTF of our decoder in a CART system is $0.5$.\ Despite the discrepancy, we believe that in terms of an efficient implementation there can still be room for improvement in a future work.\
\vspace{-4mm}
\section{Experimental Results}
\vspace{-3mm}	
The effect of label smoothing and Specaugment on both standard and factored hybrid are compared in \cref{tab:all}.\ It is possible to see that the standard hybrid improves only by 2\%, whereas the factored hybrid trained with the same alignment benefits up to 5\% relative under similar conditions, and 7\% relative with one realignment step.\ The factored model using GMM-Mono alignment has similar performance to standard CART based hybrid, making CART an optional feature for hybrid models.\ The results in \cref{tab:all} show that higher n-gram context and realignment consistently give improvement in all experiments.\ In \cref{tab:libri}, we present our baseline for the LibriSpeech 960h task.\ The experiments did not use the two mentioned regularization methods.\ However, we can see similar behavior to our SWB baseline, where in absence of label smoothing and Specaugment the quality of the initial alignment decides for the final performance.\ 

  \begin{table}[t]
  
  	\setlength{\tabcolsep}{0.4em}\renewcommand{\arraystretch}{1.1}  
  	\centering  \footnotesize 	
  	
  	\caption{The effect of Specaugment~(Specaug), label smoothing~(LS) and realignment~(Realign) in SWB 300h for factored hybrid models compared to the standard CART based model, evaluated on Hub5'00 with 4-gram LM.\ }
  	\vspace{-2mm}\label{tab:all} \footnotesize	
  	\begin{tabular}{|c|c|c|c|c|c|c|c|} 
  		\hline			
  		\multirow{2}{*}{\textbf{Specaug}}&\multirow{2}{*}{\textbf{LS}} & \multirow{2}{*}{\textbf{Model}}& \multirow{2}{*}{\textbf{Alignment}} & \multirow{2}{*}{\textbf{Realign}}& \multicolumn{3}{c|}{\textbf{[\%WER]}}  \\ \cline{6-8}
  		&                 &                     &                      &                  &  mono   & di     & tri   \\\hline
  		\multirow{3}{*}{no}&  \multirow{3}{*}{no} & CART     & \multirow{2}{*}{Tandem}  &  \multirow{3}{*}{no}&  17.3 & 15.0 & 13.4 \\ \cline{3-3} \cline{6-8}
  		&                      &\multirow{2}{*}{FH}&                               &                    &  15.1 & 14.0 & 13.5 \\ \cline{4-4} \cline{6-8}
  		&                      &                   & GMM-Mono                 &                    &  17.1 & 15.1 & 14.5 \\  \hline\hline
  		\multirow{5}{*}{no}&  \multirow{5}{*}{yes}& CART              & \multirow{3}{*}{Tandem}  & \multirow{2}{*}{no}& \multicolumn{2}{c|}{-} & 13.7 \\\cline{3-3} \cline{6-8}
  		&                      &\multirow{4}{*}{FH} &                            &                    &  15.0 & 14.0 & 13.4 \\ \cline{5-8}
  		&                               &                    &                      &         yes        &  14.6 & 13.6 & 13.1 \\ \cline{4-8}
  		&                               &                    &  GMM-Mono &         yes        &  16.1 & 14.5 & 13.8 \\ \hline\hline
  		\multirow{5}{*}{yes}&  \multirow{5}{*}{yes}& CART              & \multirow{3}{*}{Tandem}  & \multirow{2}{*}{no}& \multicolumn{2}{c|}{-} & 13.1 \\ \cline{6-8}\cline{3-3}
  		&                      &\multirow{4}{*}{FH} &                               &                    &  \textbf{14.1} & 13.3 & 12.8 \\ \cline{5-8}
  		&                               &                    &                      &         yes        & \multirow{3}{*}{-}  & \textbf{12.8} & \textbf{12.5} \\ \cline{4-5} \cline{7-8}
  		&                      &                   & \multirow{2}{*}{GMM-Mono}&         no         &          & 14.2 & 14.0 \\ \cline{5-5} \cline{7-8}
  		&                               &                    &                      &         yes        &                            & 13.7 & 13.0  \\  \hline 		
  	\end{tabular} 
  	\vspace{-6mm}	
  \end{table}
  
 Even though the focus of our work is on the elimination of state-tying, we did not limit ourselves to mere comparison of our approach to the standard hybrid.\ We also trained an enhanced model using speaker adaptation with i-vectors~\cite{kitza2019cumulative}, and in \cref{tab:literaure} we compare our Word Error Rates using both 4-gram and LSTM LMs with other state-of-the-art results in the literature.\ It is possible to see that our approach outperforms the hybrid model that is in addition using sequence discriminative training~(sMBR)~\cite{kitza2019cumulative} on averaged HUB5'00, by $1.9\%$ relative.\ The flat-start hybrid model using no state-tying and trained with lattice-free maximum mutual information~(LF-MMI)~\cite{endtoendpovey} shows also a significant weaker performance.\ Compared to the phoneme based transducer~\cite{wei} that shares many similarities to our factored hybrid in terms of model size and approach, even starting with a GMM-Mono alignment leads to $4.3\%$ relative improvement on both Hub5'0\{0,1\}.\ This gap is increased to relative $8.7\%$ and $6.1\%$ on the mentioned evaluation sets by using a better initial alignment.\ We are still behind the RNN-T model that uses additional regularization techniques~\cite{tuskernnt}.\ However, we reach a competitive result on at least two subsets of HUB5'01, by carrying out a Transformer rescoring.\ Finally, even with an LSTM LM, we perform in the same range as the attention encoder-decoder~(AED) model with Specaugment, trained for many more epochs.\
 
  \begin{table}[t]
	
	\setlength{\tabcolsep}{0.4em}\renewcommand{\arraystretch}{1.1}  
	\centering \footnotesize
	\caption{Our baseline results with Librispeech 960h and evaluated on both dev and test sets.\ }
	\vspace{-2mm}	
	
	\label{tab:libri}
		
	\begin{tabular}{|c|c|c|c|c|c|} 
		\hline		
		\multirow{2}{*}{\textbf{Model}} &\multirow{2}{*}{\textbf{Alignment}} & \multicolumn{4}{c|}{\textbf{[\%WER]}} \\ \cline{3-6}
		&        				 & dev-clean & dev-other & test-clean & test-other \\ \hline
		CART               & \multirow{2}{*}{GMM-Tri}&     4.1       & \textbf{9.3} & \textbf{4.3} &    9.9           \\ \cline{1-1} \cline{3-6}
		\multirow{2}{*}{FH}&         				 &	\textbf{3.8} &  9.4	        & \textbf{4.3} &  \textbf{9.4} \\ \cline{2-6}
		&    GMM-Mono             &		4.3      &		10.2	&	4.9		   &      10.7       \\ \hline
		
	\end{tabular}  
		
\end{table}

\begin{table}[t]
	
	\setlength{\tabcolsep}{0.08em}\renewcommand{\arraystretch}{1.2}  
	\centering \footnotesize
	\caption{Evaluation of SWB300h systems on different subsets of HUB5'00 and HUB5'01 from the literature.\  }
	\vspace{-2mm}	
	
	\label{tab:literaure}
	\begin{tabular}{|c|c|c|c||c|c||c|c|c|} 
		\hline
		
		\multirow{2}{*}{\textbf{Work}} &\multirow{2}{*}{\textbf{Model}} &\multirow{2}{*}{\textbf{\#Eps}}&\multirow{2}{*}{\textbf{LM}} &\multicolumn{2}{c||}{\textbf{HUB5'00}} &\multicolumn{3}{c|}{\textbf{HUB5'01}} \\ \cline{5-9}
		&                    & &        &  SWB & CH  & SWB & SWB2 & SWBC  \\ \hline \hline
		\cite{park2019specaugment} & AED &760& \multirow{2}{*}{RNN}                    & 6.8   & 14.1 & \multicolumn{3}{c|}{} \\\cline{1-3} \cline{5-6}
		\cite{endtoendpovey}       & Hybrid (LF-MMI)               &4&  & 9.3   & 18.9 &  \multicolumn{3}{c|}{\multirow{1}{*}{-}}\\  \cline{1-6} 
		\cite{kitza2019cumulative} & Hybrid+ivec+sMBR & - & \multirow{3}{*}{LSTM} &6.7 & 14.7 &  \multicolumn{3}{c|}{} \\ \cline{1-2} \cline{3-3}\cline{5-9}
		\cite{tuskernnt}           & RNN-T                &\multirow{2}{*}{100}& & 6.3   & 13.1 & 7.1 & 9.4 & 13.6  \\ \cline{1-2} \cline{5-9}
		\cite{wei}                 & Phon-Trans           &&                     & 7.6   & 15.4 &  8.4 & 10.6 & 15.2   \\ \hline \hline 
		\multirow{4}{*}{This}      & \multirow{2}{*}{FH+ivec}      &\multirow{3}{*}{103$^*$}&   Trafo &  6.5  & 13.7 & 7.1   &   9.5  & 13.3\\ \cline{4-9}   
		\multirow{4}{*}{Work}      &    \multirow{2}{*}{(Tandem)}  &&   LSTM & 6.8   & 14.2 &  7.5  &  10.0   & 13.8\\ \cline{4-9}   
		&      && 4-gram                              & 8.0   & 16.3 &  8.7 & 12.0  & 16.1      \\  \cline{2-9}   
								   & FH+ivec &\multirow{2}{*}{101$^*$}& LSTM                                & 7.2   & 14.8 &  7.8  &  10.7& 14.4  \\  \cline{4-9}   
		&      (GMM-Mono)                    && 4-gram                              & 8.5   & 17.1 &  9.0 & 12.3  & 16.5      \\ \hline  		
	\end{tabular} 
	\\ $*$ Phonetic training: 20 (monophone) + 53 (diphone) + 28 or 30 (triphone)\vspace{-0.3cm}
\vspace{-0.2cm}		
\end{table}

\vspace{-0.5cm}
\section{Conclusions}
\vspace{-0.3cm}
\label{sec:conclude}
In this paper, we have shown that the elimination of explicit state-tying from hybrid HMM models results in performance gains across multiple tasks.\ Our factored hybrid approach works better in combination with popular regularization techniques compared to a hybrid system using state-tying.\ This work also demonstrates that the hybrid approach is still competitive with nowadays popular sequence to sequence models.\  

\vspace{-0.3cm}
\section{Acknowledgments}
\vspace{-0.1cm} \scriptsize
This work was partly funded by the Google Faculty Research Award for "Label Context Modeling in Automatic Speech Recognition".\ This work was partially supported by the project HYKIST funded by the German Federal Ministry of Health on the basis of a decision of the German Federal Parliament (Bundestag) under funding ID ZMVI1-2520DAT04A.\ We would like to thank Christoph M. L\"uscher and Alexander Gerstenberger for providing us the LibriSpeech CART baseline and performing the lattice rescoring experiment with Transformer, respectively.\

\newpage


\bibliographystyle{IEEEbib}
{\scriptsize
\bibliography{strings,refs}}

\end{document}